\documentclass{iopart}
\usepackage[dvips]{graphicx}
\usepackage{dcolumn}
\usepackage{bm}
 \usepackage{latexsym}
 \usepackage{iopams}
 \usepackage{amsthm}
 \usepackage{amsfonts}

 \usepackage{epsfig}
 \usepackage{graphics}

\begin{document}

\title[Anisotropic universal conductance fluctuations in disordered quantum wires...]{Anisotropic universal conductance fluctuations in disordered quantum wires with Rashba and Dresselhaus spin-orbit interaction and applied in-plane magnetic field}

\author{Matthias Scheid$^{1,2}$, \.{I}nan\c{c} Adagideli$^{1}$, Junsaku Nitta$^{2}$, Klaus Richter$^{1}$}
\address{$^1$ Institut f\"ur Theoretische Physik, Universit\"at Regensburg, 93040 Regensburg, Germany}
\address{$^2$ Graduate School of Engineering, Tohoku University, 6-6-02 Aramaki-Aza Aoba, Aoba-ku, Sendai 980-8579, Japan}
\ead{Matthias.Scheid@physik.uni-r.de}

\begin{abstract}
We investigate the transport properties of narrow quantum wires realized in disordered two-dimensional electron gases in the presence of $k$-linear Rashba and Dresselhaus spin-orbit interaction (SOI), and an applied in-plane magnetic field. Building on previous work [Scheid, et al., PRL \textbf{101}, 266401 (2008)], we find that in addition to the conductance, the universal conductance fluctuations also feature anisotropy with respect to the magnetic field direction. This anisotropy can be explained solely from the symmetries exhibited by the Hamiltonian as well as the relative strengths of the Rashba and Dresselhaus spin orbit interaction and thus can be utilized to detect this ratio from purely electrical measurements.
\end{abstract}
\section{Introduction}

The ongoing miniaturization of devices based on conventional electronics is expected to hit critical boundaries soon. To be able to further improve their performance there is a growing need of adding new functionalities or to replace segments of conventional electronics devices by new building blocks. One promising candidate that could achieve either of both is the growing field of spintronics, which has seen rapid progress over the last few years~\cite{Awschalom2007}. 
Spintronics follows the vision to develop device building blocks operating on the basis of information encoded in the electron spin degree of freedom; a by now classic example is the spin-field effect transistor~\cite{Datta1990}. 
Recently much effort has been devoted to make use of spin-orbit interaction (SOI) to efficiently control the spin of the electrons. In two-dimensional electron gases (2DEGs) formed in III-V zinc-blende semiconductor heterostructures, two kinds of SOI are dominant, namely, Rashba SOI due to structural inversion asymmetry~\cite{Rashba1960} and Dresselhaus SOI due to bulk inversion asymmetry of the semiconductor crystal~\cite{Dresselhaus1955}.\\
Building on the results from Ref.~\cite{Scheid2008} in this article we present a systematic study of quantum transport in wires realized in 2DEGs with Rashba and Dresselhaus SOI subject to an additional in-plane magnetic field whose interplay gives rise to interesting effects on the transport properties~\cite{Malshukov1997,Schliemann2003,Cartoixa2003}. To this end we first introduce the system and the methods which we use for the numerical calculations in section 2. Then, in section 3, we consider as a first step the influence of the interplay of SOI and a magnetic field on the conductance and the universal conductance fluctuations (UCFs) of a purely one-dimensional quantum wire. We analyze the symmetries of the respective Hamiltonian and numerically check the predictions from random matrix theory (RMT) for the conductance and the UCFs. In particular the size of the UCFs depends on the angle between the direction of the wire and the in-plane magnetic field. As a next step in section 4 we relax the constraint of a purely one-dimensional quantum wire and investigate the transport properties of wires supporting several transversal modes. We find that for wires, whose width is much smaller than the spin precession length due to SOI, the behaviour is similar to the one-dimensional case. Before concluding with a summary of the presented results, in section 5 we put forward a detection mechanism for the relative strength of Rashba and Dresselhaus SOI based on the angular anisotropy of the transport properties found in the preceeding sections.

\section{System of choice and numerical methods}
We investigate the transport properties of a disordered quantum wire realized in a 2DEG ($x$-$y$ plane), with SOI subject to an in-plane magnetic field. For quantum wells grown in [001]-direction of III-V semiconductors there are typically two main contributions to the SOI: First the Rashba SOI~\cite{Rashba1960} given by the Hamiltonian
\begin{displaymath}
 H_\mathrm{R}
=\frac{\alpha}{\hbar} \big( \sigma_x p_y - \sigma _y p_x\big) \, .
\end{displaymath}
Second, the Dresselhaus SOI~\cite{Dresselhaus1955} described by the Hamiltonian~\cite{Lusakowski2003}:
\begin{eqnarray*}
H_\mathrm{D}=\frac{\beta}{\hbar} \Big[ &&( \sigma_x\cos{2\phi} -\sigma_y\sin{2\phi}) p_x -\\
&&(\sigma_x\sin{2\phi} +\sigma_y\cos{2\phi}) p_y \Big] \, .\nonumber
\end{eqnarray*}
Here $\phi$ denotes the angle between the $x$-direction and the $[100]$-direction of the zinc-blende crystal and $\alpha$ ($\beta$) sets the strength of the Rashba (Dresselhaus) SOI. We note that $\alpha$ can be changed by adjusting the gate voltage ~\cite{Nitta1997}, in contrast to $\beta$ which is a property of the material. Taking into account both mentioned SOI contributions and an in-plane magnetic field, the single particle Hamiltonian of a disordered quantum wire oriented in $x$-direction reads
\begin{eqnarray}\label{Ham}
H &=& \frac{p_x^2+p_y^2}{2m^*} + V_\mathrm{conf}(y) + V_\mathrm{dis}(x,y) \\
&&+ H_\mathrm{R} + H_\mathrm{D} +
\frac{\mu _\mathrm{B}g^*}{2}\vec{B}_{||}\cdot\vec{\sigma}.\nonumber
\end{eqnarray}
\begin{figure}[h!]
	\centering 
	\includegraphics[width=0.5\linewidth]{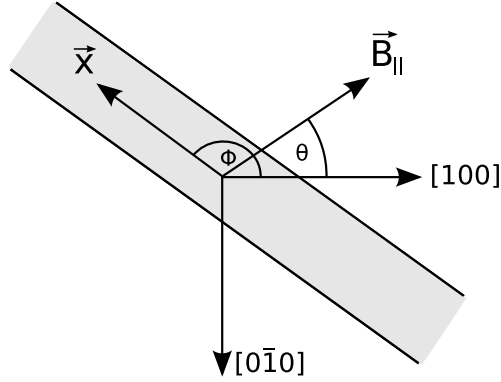}
	\caption{Relative orientation of the quantum wire (pointing in $x$-direction), the external magnetic field $\vec{B}_{||}$ and the underlying crystal lattice.}
\label{Axis}
\end{figure}
In Eq.~(\ref{Ham}) the in-plane magnetic field, whose orbital effect we neglect, is given by
\begin{equation}\label{B||}
\vec{B}_{||}=B_{||} \left[ \cos (\theta -\phi ) \hat e_x + \sin (\theta -\phi ) \hat e_y \right],
\end{equation}
where $\theta$ is the angle between the magnetic field and the $[100]$-direction of the crystal. The relative orientation of the wire, the magnetic field and the crystal is visualized in Fig.~\ref{Axis}. For later convenience we rewrite $H_\mathrm{R} + H_\mathrm{D}$ in the form of an effective magnetic field,
\begin{eqnarray*}
\vec{B}_\mathrm{SO} (\vec{p}) = \frac{2}{\mu _\mathrm{B}g^*\hbar}\bigg\{ &\Big[ &\hat e_x \beta\cos{2\phi} - \hat e_y (\alpha +\beta \sin{2\phi}) \Big] p_x +\\
&\Big[ &\hat e_x (\alpha - \beta\sin{2\phi}) - \hat e_y \beta \cos{2\phi} \Big] p_y
\bigg\}.
\end{eqnarray*}
Furthermore, $V_\mathrm{conf}(y)$ in Eq.~(\ref{Ham}) denotes the hard-wall confining potential for the quantum wire and $V_\mathrm{dis}(x,y)$ the disorder potential. We assume that the disorder is only present in the central region of the quantum wire, where transport is assumed to be fully coherent. Phase-breaking occurs in the two electronic reservoirs to which the central region of length $L$ is connected by ideal, disorder-free leads. Therefore, the length $L$ of the disordered, phase coherent region can be interpreted as a measure for the phase coherence length in the numerical calculations~\cite{Scheid2008,Schapers2006}. For those we use a discretized version of the Hamiltonian~(\ref{Ham}) that allows us to evaluate the relevant transport properties of the quantum wire by computing lattice Green's functions. For details on the method used, see e.g. Ref.~\cite{Wimmer2008}.\\
We denote the dimensionless parameters of the numerical calculations by a bar. Real physical quantities are related to these  parameters as follows: For a square lattice with spacing $a$, the dimensionless energy $\bar{E}$ is related to the real physical energy $E$ by $\bar{E}=(2m^*a^2/\hbar ^2)E$, the magnetic field is $\bar{B}=(\mu _\mathrm{B}g^*m^*a^2/\hbar ^2)B$, and the dimensionless SOI strengths are $\bar{\alpha}=(m^*a/\hbar ^2)\alpha$ and $\bar{\beta}=(m^*a/\hbar ^2)\beta$. In our numerical calculations we focus on local Anderson disorder, i.e, we choose a random onsite potential value $\bar{V}_\mathrm{dis} \in [-\bar{V}_0/2;\bar{V}_0/2]$ at each lattice point, with $\bar{V}_0$ characterizing the strength of the disorder. The elastic mean free path for the electrons in the system is then given by $l=48a\sqrt{\bar{E_\mathrm{F}}}/\bar{V}_0^2$, where $\bar{E}_\mathrm{F}$ is the scaled Fermi energy.\\
Using the recursive Green's functions method we numerically calculate the total transmission probability $T(E)$ of the quantum wire, whose value at the Fermi energy $E_\mathrm{F}$ yields the conductance of a single disorder configuration in linear response in the Landauer approach: $G=G_0 T(E_\mathrm{F})$ with $G_0=e^2/h$. The transport properties we are interested in are on the one hand the conductance $\langle G\rangle$ averaged over $N_d$ different disorder configurations and the size of the UCFs in terms of $\mathrm{var}\, G=\langle G^2\rangle - \langle G\rangle ^2$.
\section{One-dimensional quantum wire}\label{Sec:1D}
In order to understand the SOI effects appearing in narrow quantum wires of finite width~\cite{Schapers2006,Lehnen2007} (i.e. wires with several open transversal modes in the disorder free limit), it is instructive to first consider a toy model: a purely one-dimensional (1D) wire, obtained from Eq.~(\ref{Ham}) by setting $p_y=0$:
\begin{equation}\label{Ham1D}
H_\mathrm{1D} = \frac{p_x^2}{2m^*} + V_\mathrm{dis}(x) + \frac{\mu _\mathrm{B}g^*}{2}(\vec{B}_\mathrm{SO}(p_y=0) + \vec{B}_{||}) \cdot \vec{\sigma}.
\end{equation}
\begin{table}
\caption{\label{table}Symmetries of $H_\mathrm{1D}$, Eq.~(\ref{Ham1D}). The normalized averaged conductance $\langle G\rangle /G_0$ and var$(G/G_0)$ were determined for a one-dimensional disordered quantum wire with Rashba SOI and fixed $L=40a$, $\bar{V}_0=0.9$, $\phi=0$, $\bar{E}_\mathrm{F}=0.5$, $N_d=100000$, $\bar{B}_{||}=0.1$ [cases (c)-(e)], $\bar{\alpha} =0.05\sqrt{2}$ [cases (b),(d),(e)] and $\bar{\beta}=0 $.}
\begin{tabular}{ccccc}
\br
Case & Description & Symmetry class & $\langle G\rangle /G_0$ & var$(G/G_0)$\\
\mr
(a)& $\vec{B}_{||}=0$, $\vec{B}_\mathrm{SO}= 0$ & orthogonal & $0.682$ & $0.295$\\
\mr
(b)& $\vec{B}_{||}=0$, $\vec{B}_\mathrm{SO}\neq 0$ & orthogonal & $0.689$ & $0.296$\\
\mr
(c)& $\vec{B}_{||}\neq 0$, $\vec{B}_\mathrm{SO}= 0$ & orthogonal & $0.672$ & $0.148$\\
\mr
(d)& $\vec{B}_{||},\vec{B}_\mathrm{SO}\neq 0$, $\vec{B}_{||}\parallel\vec{B}_\mathrm{SO}$ & orthogonal & $0.679$ & $0.148$\\
\mr
(e)& $\vec{B}_{||},\vec{B}_\mathrm{SO}\neq 0$, $\vec{B}_{||}\nparallel\vec{B}_\mathrm{SO}$ & unitary & $\begin{array}{c}>0.679 \\ <0.720\end{array}$ & $\begin{array}{c}>0.105 \\ <0.148 \end{array}$\\
\br
\end{tabular}
\end{table}
In the following we consider various situations, where the SOI and/or the in-plane magnetic field are either absent or sufficiently strong that the relevant spin rotation time is shorter than the escape time or the dephasing time and therefore are strong enough to change the symmetry class. Depending on the presence of $\vec{B}_\mathrm{SO}$ and $\vec{B}_{||}$ and on their relative angle, $H_\mathrm{1D}$ in Eq.~(\ref{Ham1D}) belongs to different symmetry classes as summarized in Table~\ref{table}.\\
Apart from the trivial case (a), where SOI and external magnetic fields are absent, spin is a good quantum number of $H_\mathrm{1D}$ also for the cases (b)-(d). Although the full Hamiltonian $H_\mathrm{1D}$ does not possess time-reversal symmetry, $\mathcal{T}^{-1}H_\mathrm{1D}\mathcal{T}\neq H_\mathrm{1D}$, seperate Hamiltonians for spin up and spin down, which are both time-reversal symmetric (for cases (b) and (d) after an additional gauge transformation), can be written down, since they decouple:
\begin{eqnarray}\label{HamDiag}
H_\mathrm{1D}=\left( \begin{array}{cc} 
H_+ & 0 \\
0 & H_-
\end{array} \right),\\
\mathrm{with}\quad H_{\pm}= \frac{p_x^2}{2m^*} + V_\mathrm{dis}(x) \pm 
\frac{\mu _\mathrm{B}g^*}{2}\left(|\vec{B}_\mathrm{SO}(p_y =0)| + B_{||}\right).\nonumber
\end{eqnarray}
Due to their invariance under complex conjugation the Hamiltonians $H_\pm$ (for cases (b) and (d) after an additional gauge transformation) belong to the orthogonal symmetry class, and obviously no mixing of the spins occurs. In case (e) however, the spin degrees of freedom mix and time reversal symmetry is broken. As a consequence $H_\mathrm{1D}$ possesses only unitary symmetry.\\
Those symmetries have important consequences for the transport properties of the disordered quantum wire. In RMT developed for coherent quantum transport~\cite{Beenakker1997} it was shown that systems with unitary, orthogonal and symplectic symmetry exhibit different quantum corrections to the conductance and different magnitudes of the UCFs~\cite{Mello1988,Mello1991}. Although RMT gives quantitative predictions only for quantum wires with a large number of transversal modes, it can help us to understand the qualitative behaviour of both the average conductance and the UCFs for the one-dimensional quantum wire treated in this section.\\
In the cases (a)-(d) the values of $\langle G\rangle$ are similar (showing weak localization, i.e. a negative correction to the classically expected conductance), while only the averaged conductance in case (e) is increased, both due to the mixing of the spins and due to the absence of orthogonal symmetry.
Since the physical systems of cases (a)-(d) are different, also the values of $\langle G\rangle$ differ slightly, since we cannot compute the quantum interference corrections to the conductance directly but only the total conductance.\\
Considering the numerical values for the UCFs presented in Table~\ref{table} we observe that, although cases (a)-(d) have orthogonal symmetry, for (a) and (b) var$(G/G_0)$ has twice the value than for (c) and (d). This can be understood by considering correlations between $H_+$ and $H_-$ from Eq.~(\ref{HamDiag}). In cases (a) and (b) $H_+$ and $H_-$ are the time-reversed of each other ($H_+^*=H_-$), resulting in double the value for the UCFs~\cite{Lyanda-Geller1994} compared to cases (c) and (d), where due to the Zeeman splitting the two separate Hamiltonians are uncorrelated, i.e. $H_+^*\neq H_-$. On the other hand for case (e) var$(G/G_0)$ is even smaller than for (a)-(d), since the system then mixes spins and has unitary symmetry.\\
\begin{figure}[h!]
	\centering 
	\includegraphics[width=0.5\linewidth]{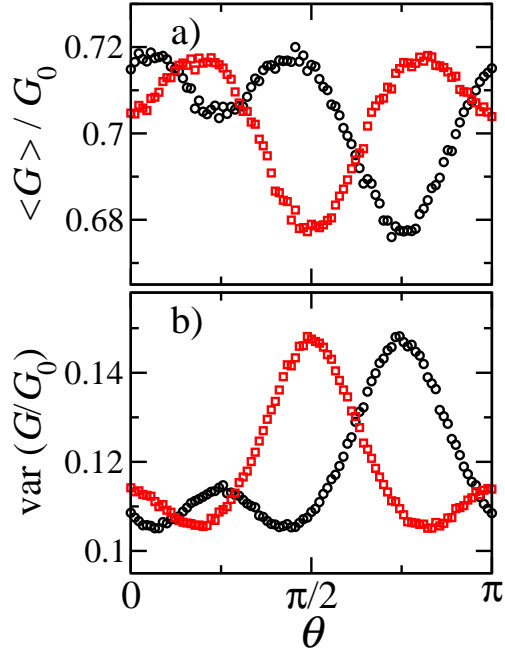}
	\caption{(Color online) Conductance of a one-dimensional quantum wire with fixed $L=40a$, $\bar{V}_0=0.9$, $\phi=0$, $\bar{E}_\mathrm{F}=0.5$, $\bar{B}_{||}=0.1$ with either $\bar{\alpha} =\bar{\beta}=0.05$ (black circles) or $\bar{\alpha} =0.05\sqrt{2}$ and $\bar{\beta}=0 $ (red squares). Panel a) and b) display the averaged conductance and UCFs, respectively, as a fuction of the magnetic field direction $\theta$. For all curves averaging over $N_d=100000$ disorder configurations was performed. Since, owing to the Onsager relations, all the effects observed in the range $\theta\in [0;\pi ]$ naturally repeat themselves at an angle $\theta +\pi$ in the range $\theta\in [\pi ;2\pi ]$, we restrict the presentation of $\langle G \rangle$ and $\mathrm{var}(G/G_0)$ to $\theta\in [0;\pi ]$.}
\label{Fig1D}
\end{figure}
In Fig.~\ref{Fig1D} we present numerical calculations for the averaged conductance of the wire and the UCFs as a function of the angle of the magnetic field $\theta$ (see Eq.~\ref{B||}) at finite $\vec{B}_{||}$ and $\vec{B}_\mathrm{SO}$ for the case of equal Rashba and Dresselhaus SOI strength (black circles) and pure Rashba SOI (red squares). In the upper panel we see two pronounced dips for each conductance curve, separated by an angle $\pi /2$. The position of the absolute minimum can be identified as the angle of the magnetic field where $\vec{B}_{||} \parallel\vec{B}_\mathrm{SO}$, corresponding to case (d) of Tab.~\ref{table}. There, due to the orthogonal symmetry, we observe weak localization, while for all other angles [case (e)] spin is not a good quantum number, and the negative conductance correction is reduced. 
The less pronounced dip corresponding to $\vec{B}_{||} \perp\vec{B}_\mathrm{SO}$ is due to an antiunitary symmetry exhibited by $H_\mathrm{1D}$ at this angle. Although the total Hamiltonian $H_\mathrm{1D}$ obviously mixes spins and thus cannot be written in block-diagonal form as for case (d) (see Eq.~\ref{HamDiag}) it nevertheless possesses orthogonal symmetry: $\mathcal{C}^{-1}\tilde{H}_\mathrm{1D}\mathcal{C}=\tilde{H}_\mathrm{1D}$, resulting in weak localization. There $\tilde{H}_\mathrm{1D}$ is the spin rotated Hamiltonian where $\vec{B}_{||}$ is aligned along the $x$-direction, and $\mathcal{C}$ is the operator of complex conjugation. This orthogonal symmetry with nonconserved spin yields a value for $\langle G\rangle$ that is in-between the values of the orthogonal case (d) with conserved spin and the unitary case (e), where spins mix.\\
For the UCFs shown in the lower panel of Fig.~\ref{Fig1D} we observe peaks at the very same positions as the dips in the conductance. This is in line with the expectations from RMT predicting a larger magnitude of the UCFs for an orthogonal ensemble such as in case (d) than for a unitary one as in case (e)~\cite{Lyanda-Geller1994,Aleiner2001}.\\
The dips/peaks in the averaged conductance/UCFs are most pronounced, when $B_{||}$ and $|\vec{B}_\mathrm{SO}|$ are of comparable strength, because then spin mixing, i.e. spin relaxation, is strongest. For $B_{||}$ much smaller/larger than $|\vec{B}_\mathrm{SO}|$ the total magnetic field $\vec{B}_{||}+\vec{B}_\mathrm{SO}$ is strongly aligned along $\vec{B}_\mathrm{SO}$ or $\vec{B}_{||}$, respectively, reducing the mixing of the spins. Then, e.g., weak localization~\cite{Scheid2008,Meijer2004,Meijer2005} is recovered.\\
The interplay of Rashba SOI, Dresselhaus SOI, an in-plane magnetic field and quantum confinement was also considered in other nanoscale systems~\cite{Bandyopadhyay2005,Dolcini2008,Nazmitdinov2008}, and the symmetry classes given in Table~\ref{table} can be identified as the origin for the angular dependence of different transport quantities found in those works.\\

\section{Quantum wire of finite width}
\begin{figure}[tb]
	\centering 
	\includegraphics[width=0.5\linewidth]{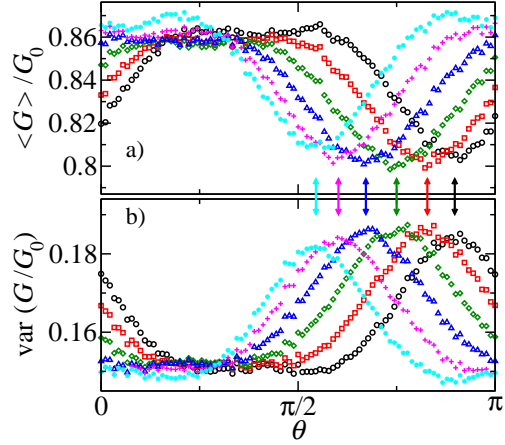}
	\caption{(Color online) Conductance of a quantum wire with fixed $W=20a$, $L=150a$, $l\approx 17.3a$, $\phi=0$, $\bar{E}_\mathrm{F}=0.5$, $\bar{B}_{||}=0.008$, $\bar{\alpha}+\bar{\beta}=0.02$ with varying ratio $\alpha /\beta=7$ (cyan stars), 3 (magenta plus), 5/3 (blue triangles), 1 (green diamonds), 3/5 (red squares), 1/3 (black circles). Panel a) and b) display the conductance $\langle G\rangle$ and UCFs, respectively, as a fuction of the magnetic field direction $\theta$. For all curves averaging over $N_d=100000$ disorder configurations was performed. The arrows indicate the respective angles, where $\vec{B}_{||} \parallel\vec{B}_\mathrm{SO}(p_y=0)$.}
\label{Fig2D}
\end{figure}

In the previous section we analyzed the transport properties of the toy model of a strictly 1D quantum wire and found interesting behaviour from the interplay of Rashba and Dresselhaus SOI and an in-plane magnetic field. In this section we extend those considerations to realistic quantum wires of finite width $W$ which allows one to make predictions for real experiments.\\
In an extended 2DEG with SOI usually a positive correction to the classically expected conductance is found, called weak antilocalization. Its appearance requires spin relaxation, which in our case is provided by the interplay between SOI and scattering at impurities, which randomizes the spin state of the electrons, namely D'yakonov-Perel' spin relaxation~\cite{Dyakonov1971}. However, it was shown, that confinement can lead to a supression of spin relaxation in quantum dots~\cite{Aleiner2001,Zaitsev2005,Zaitsev2005b}, which becomes visible as a crossover from weak antilocalization to weak localization in quantum wires upon reducing their width~\cite{Scheid2008,Schapers2006,Lehnen2007,Kettemann2007}.\\
Although, except for the special case of $\alpha =\pm\beta$, the spin is no longer a good quantum number for a disordered quantum wire of finite width, in Ref.~\cite{Scheid2008} it was shown that the angular anisotropy in the conductance of the 1D quantum wire can also be found in wires of finite width as long as $W$ is still much smaller than the spin-precession lengths $L_\mathrm{SO}^{\alpha}=\pi\hbar ^2/(m^*\alpha)$ due to Rashba SOI and $L_\mathrm{SO}^{\beta}=\pi\hbar ^2/(m^*\beta)$ due to Dresselhaus SOI.\\
Having found in the previous section that the UCFs show an equivalent anisotropy, in Fig.~\ref{Fig2D} we analyze the conductance and the UCFs for a quantum wire with four transversal modes but in the regime of $W\ll L_\mathrm{SO}^{\alpha /\beta}$. Both the averaged conductance and the UCFs show similar behaviour as their 1D counterparts in Fig.~\ref{Fig1D}. Specifically the minimum/maximum of the conductance/UCFs appears at the angle where the magnetic field is parallel to the effective magnetic field due to SOI for a $k$-vector along the wire direction $\hat{x}$: $\vec{B}_{||} \parallel\vec{B}_\mathrm{SO}(p_y=0)$. The second dip/peak at $\vec{B}_{||} \perp\vec{B}_\mathrm{SO}(p_y=0)$ that was observable in Fig.~\ref{Fig1D} for the 1D quantum wire is strongly suppressed and almost not visible anymore for the parameters chosen for the calculations of Fig.~\ref{Fig2D}.\\
Introducing the angle $\xi =\arctan \left[ \beta \cos 2\phi / (\beta\sin 2\phi -\alpha ) \right]$ we define two new Pauli matrizes by
\begin{displaymath}
\sigma _1= \cos \xi\, \sigma _x + \sin \xi\, \sigma _y,\quad\mathrm{and}\quad \sigma _2= -\sin \xi\, \sigma _x + \cos \xi\, \sigma _y .
\end{displaymath}
In order to understand the similarity of the results from a strictly 1D quantum wire and a very narrow quantum wire of finite width we perform the unitary transformation $U=\exp \left[-\mathrm{i}(\gamma /\hbar) m^* y \sigma _1\right]$ similar to that used in Ref.~\cite{Aleiner2001} on the Hamiltonian $H$ from Eq.~(\ref{Ham}):
\begin{eqnarray}\label{HamUT}
\tilde{H} &=& U^{\dagger}HU=\frac{p_x^2+p_y^2}{2m^*} + V_\mathrm{conf}(y) + V_\mathrm{dis}(x,y) -\frac{\gamma^2m^*}{2\hbar ^2}\\
&+& \frac{\mu _\mathrm{B}g^*}{2}
\vec{B}_{||}
\cdot\vec{\sigma} 
+ \frac{1}{\hbar}\Big( - \alpha\sigma _y + \beta\cos 2\phi \,\sigma _x - \beta\sin 2\phi \,\sigma _y\Big) p_x 
\nonumber \\
&+& \big[  \Delta _B + \Delta _\mathrm{SO} \big] \sigma _\delta \nonumber ,
\end{eqnarray}
Here we introduced the following parameters:
\begin{displaymath}
\gamma =\sqrt{\alpha ^2+\beta ^2 - 2 \alpha \beta \sin 2\phi},
\end{displaymath}
\begin{displaymath}
\Delta _B =\mu _\mathrm{B}g^*( B_x \sin \xi  -B_y \cos \xi  ),
\end{displaymath}
\begin{displaymath}
\Delta _\mathrm{SO}=\frac{2}{\hbar} p_x \frac{\beta ^2 -\alpha ^2}{\gamma}\quad \mathrm{and}
\end{displaymath}
\begin{displaymath}
\sigma _\delta = \sin \left(\gamma m^* y /\hbar ^2\right) \cos \left(\gamma m^* y /\hbar ^2\right) \sigma _z+ \sin ^2 \left(\gamma m^* y /\hbar ^2\right) \sigma _2 .
\end{displaymath}
When we expand the transformed Hamiltonian in lowest orders of the parameter $\gamma m^* y /\hbar ^2$ and neglect the terms of second or higher order in SOI strength, $\Delta _\mathrm{SO} \sigma _\delta$, or SOI strength times magnetic field strength, $\Delta _{B} \sigma _\delta$, only the first two lines of Eq.~(\ref{HamUT}) remain and we recover the spin part of the one-dimensional Hamiltonian from Eq.~(\ref{Ham1D}). Therefore, the toy model from section~\ref{Sec:1D} gives reasonable results also for quantum wires of finite width, as long as $W\ll \hbar ^2/ (\gamma m^*)$.
\section{All-electrical detection of the ratio $\alpha /\beta$}
\begin{figure}[h!]
	\centering 
	\includegraphics[width=0.5\linewidth]{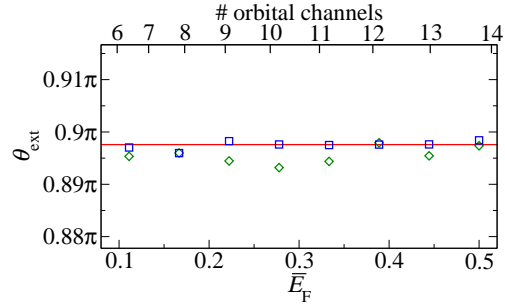}
	\caption{(Color online) Numerically determined $\theta _\mathrm{ext}$ for different values of the Fermi energy, respectively number of open channels, for a quantum wire with fixed $W=60a$, $L=450a$, $\bar{U}_0=0.785$, $\phi=\pi /2$, $\bar{B}_{||}=1/900$, $\bar{\alpha}=1/20$, $\bar{\beta}=1/60$ and $N_d=15000$. The values of $\theta _\mathrm{ext}$ determined from the minimum (maximum) of the conductance (UCFs) are presented as diamonds (squares) and compared to the expected value $\theta _\mathrm{ext}=\arctan(-\beta /\alpha )=\arctan(-1/3)\approx 0.898\pi$  from Eq.~(\ref{Detect}).}
\label{FigDetection}
\end{figure}

Since the direction of the effective magnetic field $\vec{B}_\mathrm{SO}$ depends on both $\alpha$ and $\beta$ the extremum in the conductance traces, i.e. the minimum (maximum) of the conductance (UCFs), can be used to determine the relative strength of Rashba and Dresselhaus SOI. For a $k$-vector along 
the wire the direction of $\vec{B}_\mathrm{SO}$ is given by~\cite{Scheid2008}
\begin{equation}\label{Detect}
\theta _\mathrm{ext} =\arctan\left(-\frac{\alpha\cos\phi+\beta\sin\phi}{\beta\cos\phi+\alpha\sin\phi}\right).
\end{equation}
We have seen in the previous section that, if the two requirements $W\ll L_\mathrm{SO}^{\gamma}=\pi\hbar ^2/(m^*\gamma)$ and $|\vec{B}_{||}|<|\vec{B}_\mathrm{SO}|$ are fulfilled, the minimum/maximum of the conductance/UCFs appears at $\theta _\mathrm{ext}$ from Eq.~(\ref{Detect}). In Ref.~\cite{Scheid2008} it was shown numerically that this detection mechanism can work in realistic quantum wire geometries, for a wide range of parameters. Since in a realistic experimental situation the number of transversal orbital channels is usually higher than four (the number of channels considered in Ref.~\cite{Scheid2008}), in Fig.~\ref{FigDetection} we present an extension of that work to a larger channel number. To this end we determine the minimum/maximum of the conductance/UCFs from the angular dependence of the conductance/UCFs. Comparison to the value from Eq.~(\ref{Detect}) confirms the applicability of the method also for systems with a higher electron density, i.e. higher Fermi energy or higher number of transversal orbital channels.\\
When fixing the lattice spacing $a=4$nm and using the typical values of an InAlAs/InGaAs heterostructure ($m^*=0.05m_0$, $g^*=3$), the parameters used in Fig.~\ref{FigDetection} correspond to $W=240$nm, $\alpha\approx 1.9\cdot10^{-12}$eVm, $\beta\approx 6.3\cdot10^{-13}$eVm, $B_{||}\approx 0.61$T and for the highest Fermi energy considered ($\bar{E}_\mathrm{F}=0.5$) $E_\mathrm{F}\approx 24$meV and are thus in reach of present day experiments~\cite{Meijer2004,Bergsten2006}.

\section{Summary and conclusions}
We have shown that the conductance and the UCFs of quantum wires exhibit an angular anisotropy with the direction of an in-plane magnetic field. Its exact form is determined by the strengths of Rashba and Dresselhaus SOI and the wire direction with respect to the crystal lattice.\\
Both the minimum of the conductance as well as the maximum in the UCFs can be used to determine the relative strength of Rashba and Dresselhaus SOI from a transport measurement, which is especialy appealing, since it does not rely on any fitting parameters. When confirmed experimentally it will be a valuable addition to the existing optical techniques~\cite{Ganichev2004,Averkiev2006,Meier2007,Giglberger2007}. 
\ack
The authors thank M. Wimmer for fruitful discussions. 
MS acknowledges funding from JSPS and the {\em Studienstiftung des Deutschen Volkes},
IA and KR from DFG through SFB 689 and JN from MEXT.


\end{document}